\documentclass[reprint, amsmath,amssymb, aps,]{revtex4-1}

\usepackage{graphicx}% Include figure files
\usepackage{dcolumn}% Align table columns on decimal point
\usepackage{bm}% bold math
\usepackage[utf8]{inputenc}
\usepackage{hyperref}
\begin{document}

\preprint{APS/123-QED}

\title{High Curie Temperature Ferromagnetic Semiconductor: Bimetal Transition Iodide V$_2$Cr$_2$I$_9$}% Force line breaks with \\

\author{Yulu Ren}
\affiliation{State Key Laboratory of Metastable Materials Science and Technology \& Key Laboratory for Microstructural Material Physics of Hebei Province, School of Science, Yanshan University, Qinhuangdao 066004, China}
\author{Qiaoqiao Li}
\affiliation{State Key Laboratory of Metastable Materials Science and Technology \& Key Laboratory for Microstructural Material Physics of Hebei Province, School of Science, Yanshan University, Qinhuangdao 066004, China}
\author{Wenhui Wan}
\affiliation{State Key Laboratory of Metastable Materials Science and Technology \& Key Laboratory for Microstructural Material Physics of Hebei Province, School of Science, Yanshan University, Qinhuangdao 066004, China}
\author{Yong Liu}
\affiliation{State Key Laboratory of Metastable Materials Science and Technology \& Key Laboratory for Microstructural Material Physics of Hebei Province, School of Science, Yanshan University, Qinhuangdao 066004, China}
\author{Yanfeng Ge}
\email{yfge@ysu.edu.cn}
\affiliation{State Key Laboratory of Metastable Materials Science and Technology \& Key Laboratory for Microstructural Material Physics of Hebei Province, School of Science, Yanshan University, Qinhuangdao 066004, China}

\date{\today}% It is always \today, today,

\begin{abstract}
Bimetal transition iodides in two-dimensional scale provide an interesting idea to combine a set of single-transition-metal ferromagnetic semiconductors together. Motivated by structural engineering on bilayer CrI$_3$ to tune its magnetism and works that realize ideal properties by stacking van der Waals transitional metal dichalcogenides in a certain order. Here we stack monolayer VI$_3$ onto monolayer CrI$_3$ with a middle-layer I atoms discarded to construct monolayer V$_2$Cr$_2$I$_9$. Based on this crystal model, the stable and metastable phases are determined among 7 possible phases by first-principles calculations. It is illustrated that both the two phases have Curie temperature $\sim$ 6 (4) times higher than monolayer CrI$_3$ and VI$_3$. The reason can be partly attributed to their large magnetic anisotropy energy (the maximum value reaches 412.9 $\mu$eV/atom). More importantly, the Curie temperature shows an electric field and strain dependent character and can even surpass room temperature under a moderate strain range. At last, we believe that the bimetal transition iodide V$_2$Cr$_2$I$_9$ monolayer would support potential opportunities for spintronic devices.
\end{abstract}

\maketitle

%\tableofcontents

\section{INTRODUCTION}

The intrinsic ferromagnetic semiconductors (FMSs) in atomically thin scale have exerted serious interests in spintronic devices and data storage area in recent years, when the traditional bulk materials are hard to satisfy the pursuit of nano-sized spintronic devices with high integration \cite{Huang2020, Huang2017, Dietl2014, McGuire2015, Zhong2017, MacDonald2005, Dietl2010}. Likewise, FMSs can also be used for spin injection, generation, manipulation, and detection with the help of nowadays well-developed semiconductor technology \cite{Li2016}. Different from bulk magnets, two-dimensional (2D) magnetism though has long been at the heart of numerous experimental and technological advances \cite{Hellman2017, Burch2018, Zak1990, Arnold1997}, is rare for its vulnerability to thermal energy. Therefore, ferromagnetism in 2D scale is conditional and should possess such fundamentals: (i) magnetic anisotropy, (ii) exchange interaction \cite{Gong2019, Mermin1966}.  Along with the fact that ferromagnetism and semiconductivity are not very well compatible, the amounts of FMSs in atomically thin are limited, let alone those with high Curie temperature ($T_c$) \cite{Feng2017}.

A series of ferroelectric magnets are reported with high $T_c$ observed in experiment such as layered metallic Fe$_3$GeTe$_2$ ($\sim$ 225 K) \cite{Fei2018}, VSe$_2$ ($\sim$ 300 K)  \cite{Bonilla2018} and V doped WSe$_2$ ($\sim$ 360K) \cite{Yun2020} but they all show an absence of semiconductivity. In addition, although theoretical works have predicted FMSs with high $T_c$ such as GdI$_2$ monolayer ($\sim$ 241 K) \cite{Wang2020}, CrX (X = P, As, $\sim$ 232 and 855 K) \cite{Ma2020}, zinc-blende CrC ($\sim$ 1900 K) \cite{Huang2019} and Fe$_2$Si ($\sim$ 780 K \cite{Sun2017}), the ferromagnetism in 2D FMSs can only be observed at a relatively low temperature in experiment. Such as recently synthesized layered CrI$_3$ ($\sim$ 45 K) \cite{Huang2017, PJiang2018, Soriano2019}, VI$_3$ ($\sim$ 51 K) \cite{Kong2019, Tian2019}, Cr$_2$Ge$_2$Te$_6$ ($\sim$ 67.9 K) \cite{Gong2017} and Cr$_2$Si$_2$Te$_6$ ($\sim$ 33 K) \cite{Lin2016}. So, efforts have been devoted to tune the ferromagnetism and enhance $T_c$ by electrons doping, tunable gate and stacking \cite{Deng2018, Yan2020, Frey2019, Jiang2018, Wang2018}. For example, the magnetic phase can be switched in bilayer CrI$_3$, and $T_c$ can also be elevated in monolayer CrI$_3$ by switching iron gate \cite{Huang2017, Huang2018, McGuire2015}. Whose effect is also observed in Cr$_2$Ge$_2$Te$_6$ \cite{Verzhbitskiy2020}. Besides, breaking long-range out-of-plane symmetry and employing tensile strain can also effectively affect the magnetism, according to recent works \cite{Ren2020, Leon2020, Hu2020, Cui2020, Abdollahi2020}.
%our previous prediction \cite{Ren2020}.

Inspired by the layer-number dependent magnetic phase switch in FMSs CrI$_3$ \cite{Huang2017}, we try to build a bilayer VI$_3$ model [left panel of Fig. \ref{1}(a)] and investigate magnetic behavior different with its prototype. Unfortunately, it is found that no significant change in magnetic anisotropy or $T_c$ compared with monolayer VI$_3$ \cite{supply}. Then we try to break out-of-plane symmetry by substitute one V atoms layer with Cr atoms [middle panel of Fig. \ref{1}(a)], but the calculation of phonon frequency reveals that this bilayer crystal is dynamically unstable with a large imaginary frequency. Finally, also inspired by a set of crystal models such as ferroelectrics layered In$_2$Se$_3$ \cite{Ding2017}, we peel one  I layer off to combine the bilayer compound into monolayer bimetal transition iodide V$_2$Cr$_2$I$_9$ [right panel of Fig. \ref{1}(a)]. It is rather interesting that  the imaginary frequency is eliminated after the structural alteration.  Therefore, this work is discussed mainly in three parts. Firstly, we give a systematical investigation on wither the structural alteration is theoretically feasible by using first-principles calculations to confirm the stable and metastable states form 7 different phases of monolayer V$_2$Cr$_2$I$_9$. Next, we discuss that the two feasible phases show an enhancement on ferromagnetism and higher $T_c$ compared to their prototypes. At last, efforts are devoted to further tune the $T_c$ by external electric field and strain. According to the results, V$_2$Cr$_2$I$_9$ monolayer is likely to be synthesized and shows a certain perspective in spintronic field as a member of layered van der Waals FMSs family.

\section{RESULTS AND DISCUSSION}
\begin{figure*}[htbp]
    \centering
    \includegraphics[scale=0.56]{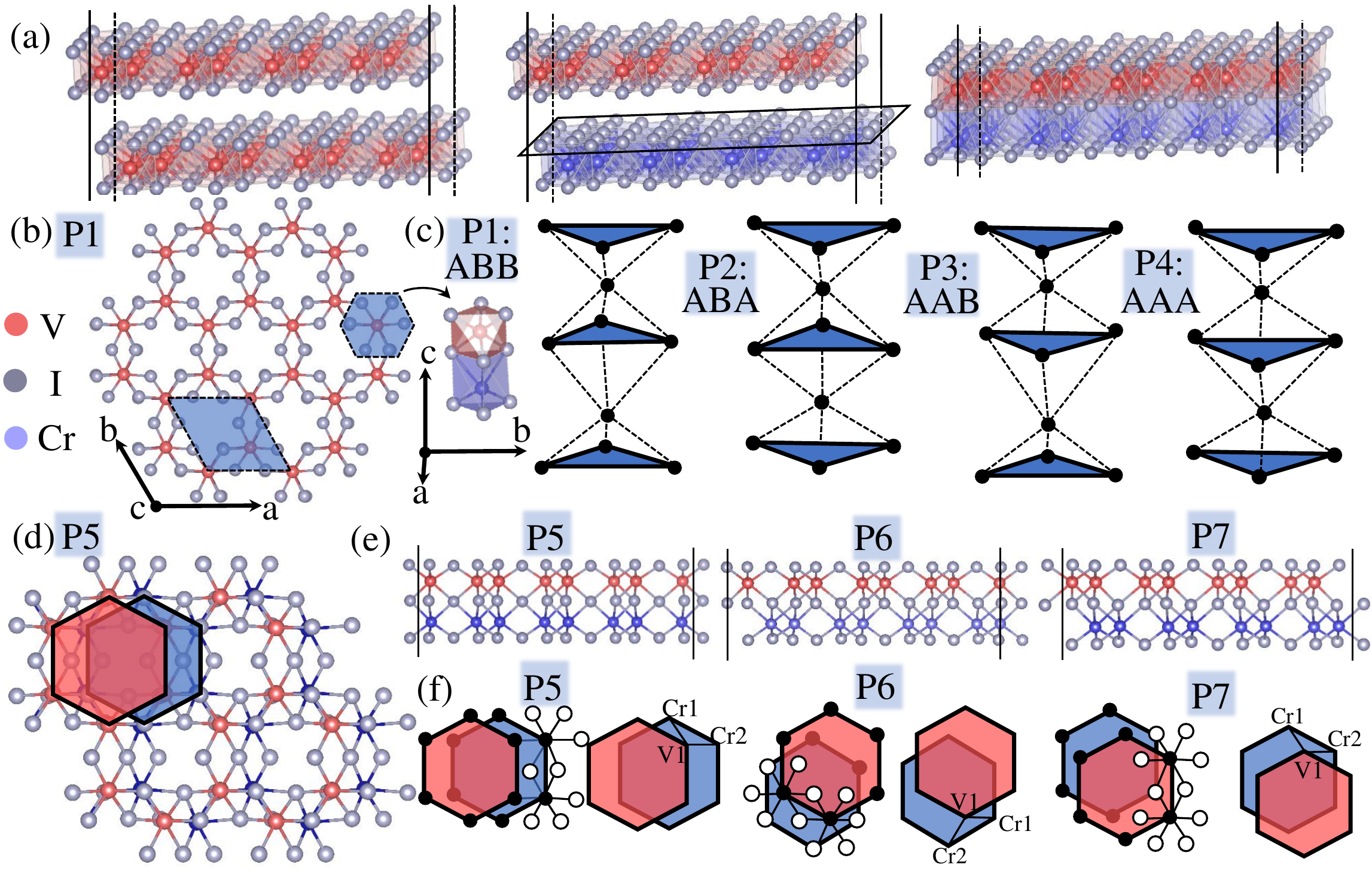}
       \caption{\label{1}(a) Structural evolution from bilayer VI$_3$ to monolayer V$_2$Cr$_2$I$_9$. (b) Top view of the optimized structure of P1. The rhomboid area donates a primitive cell containing two Cr, two V and nine I atoms, while the hexagon area denotes a V-Cr unit. Red, gray and violet balls denote V, I and Cr atom, respectively. (c) Schematic diagram for side view of a V-Cr unit for P1 $\sim$ P4. (d) Top view of optimized structure of P5, red and blue areas denote hexagons formed by V and Cr hexagons, respectively. (e) Side view of crystal structure for P5 $\sim$ P7. (f) Corresponding  top-view schematics for P5 $\sim$ P7. The solid and hollow circles denote transition metal and I atoms, respectively. Details for lattice parameters, bond lengths and angles are shown in supporting materials \cite{supply}.}
     \end{figure*}

The monolayer V$_2$Cr$_2$I$_9$ in Fig. \ref{1}(a) is obtained from monolayer CrI$_3$ (P3$_1$12, No.151) and VI$_3$ (R$\overline{3}$, No.148). The calculated lattice constants of monolayer CrI$_3$ and VI$_3$ is a$_1$ = b$_1$= 6.907 {\AA} and a$_2$ = b$_2$ = 6.824 {\AA}, respectively, consistent with the experiments \cite{McGuire2015, Tian2019}. So a 1.05 \% ($<$ 5 \%) mismatch in ab lattice constants is obtained [(a$_1$-a$_2$)/a$_1$]. Figures \ref{1} (a) and \ref{1} (b) show a honeycomb monolayer with V-Cr arraying parallel with c axial. According to different polyhedron types formed by each Cr and V atoms with their nearest six I atoms, four possible phases named P1 $\sim$ P4 are obtained as shown in Fig. \ref{1}c. In P1 model, each Cr-V unit shows an ABB stacking with a V-I$_6$ octahedron and a Cr-I$_6$ trigonal prism. Not only the discrepancy in stacking order, but a relative displacement along ab plane is likely to obtain interesting phases. So, we give the whole V layer a small rigid shift within ab plane based on ABB stacking \cite{Zollner2019}, followed by atomic relaxation.  It shows the distances between V1 and Cr1 $d_{V1-Cr1}$ are right equal to $d_{V1-Cr2}$. That is, by moving V layer of P1 to [1/3 0 0], [1/3 1/3 0] and [0 1/3 0], another three phases named P5, P6 and P7, are confirmed respectively [Figs. \ref{1}(d), \ref{1}(e) and \ref{1}(f)].

\begin{figure}[htbp]
    \centering
    \includegraphics[scale=0.52]{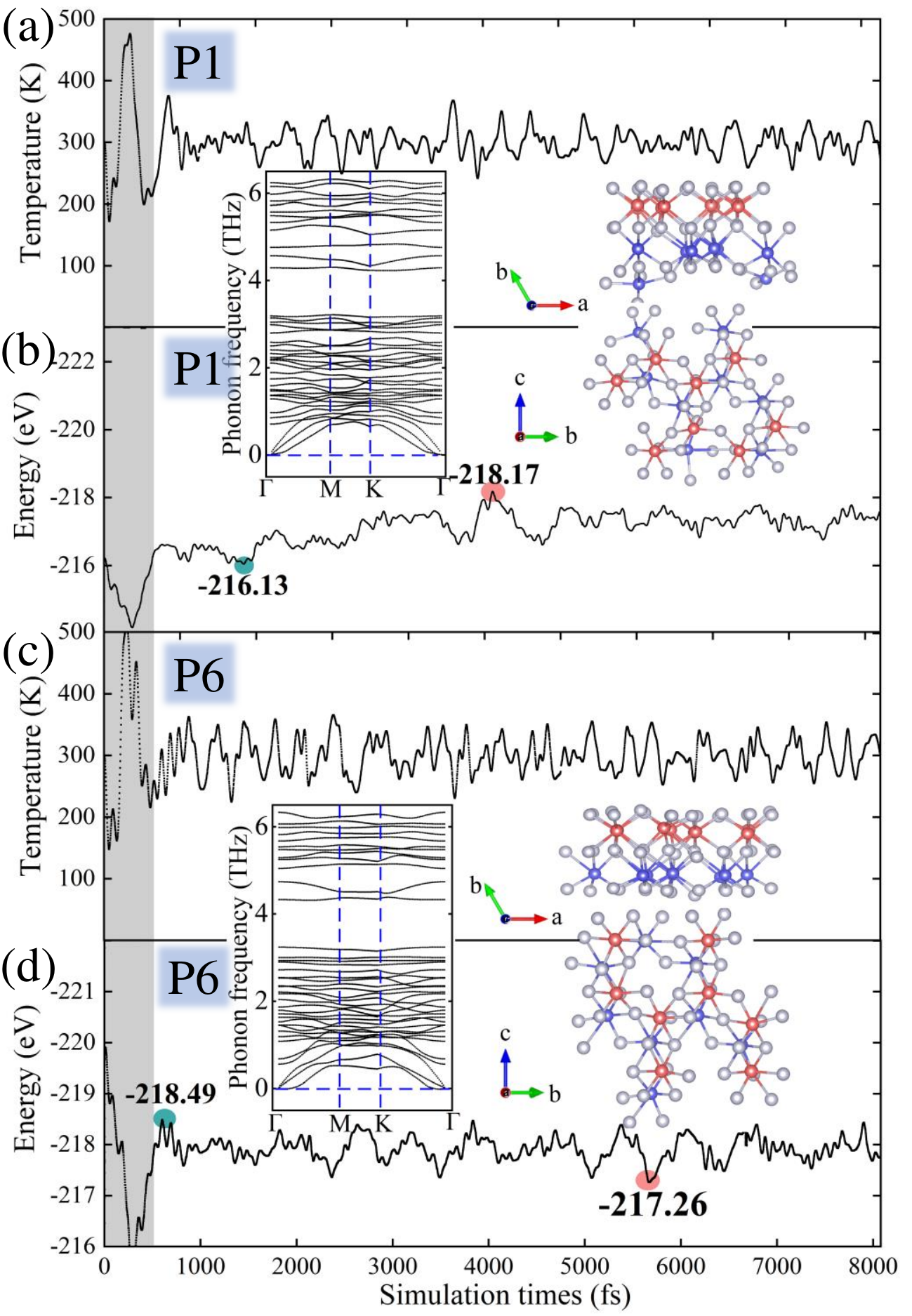}
       \caption{\label{2}(a, c) Crystal temperature as a function of time by AIMD simulations at 300 K for P1 and P6, respectively. (b, d) Total potential energy as a function of time for P1 and P6, respectively. The green and red points mark the maximum and minimum after 430 fs (grey area). The left insert figures are corresponding phonon frequency band structures along high-symmetry points, while the right inserts are corresponding top and side view of crystal structures after 8000 fs AIMD simulations. Red, blue and silver bulls denote V, Cr and I atoms, respectively.}
\end{figure}

%\subsection{Electronic Properties of VXY}
We next check the stability and further determine stable or metastable phases. Among P1 to P4, only P1 shows an absence of imaginary frequency while P2 $\sim$ P4 are dynamically unstable with imaginary frequency such as -2.815 THz in P3. Among P5 to P7, we find P6 also shows dynamic stability (Fig. \ref{2}). Furthermore, the ab initial molecular dynamic (AIMD) simulations lasting 8000 fs at room temperature are employed on P1 and P6.  The total potential energy shows a small fluctuation range in P1 and P6 and the final structures remain basically honeycomb lattice, indicating thermal stable.
%\subsection{Competitive Mechanism between FM and AFM.}

\begin{figure}[htbp]
 \centering
 \includegraphics[scale=0.17]{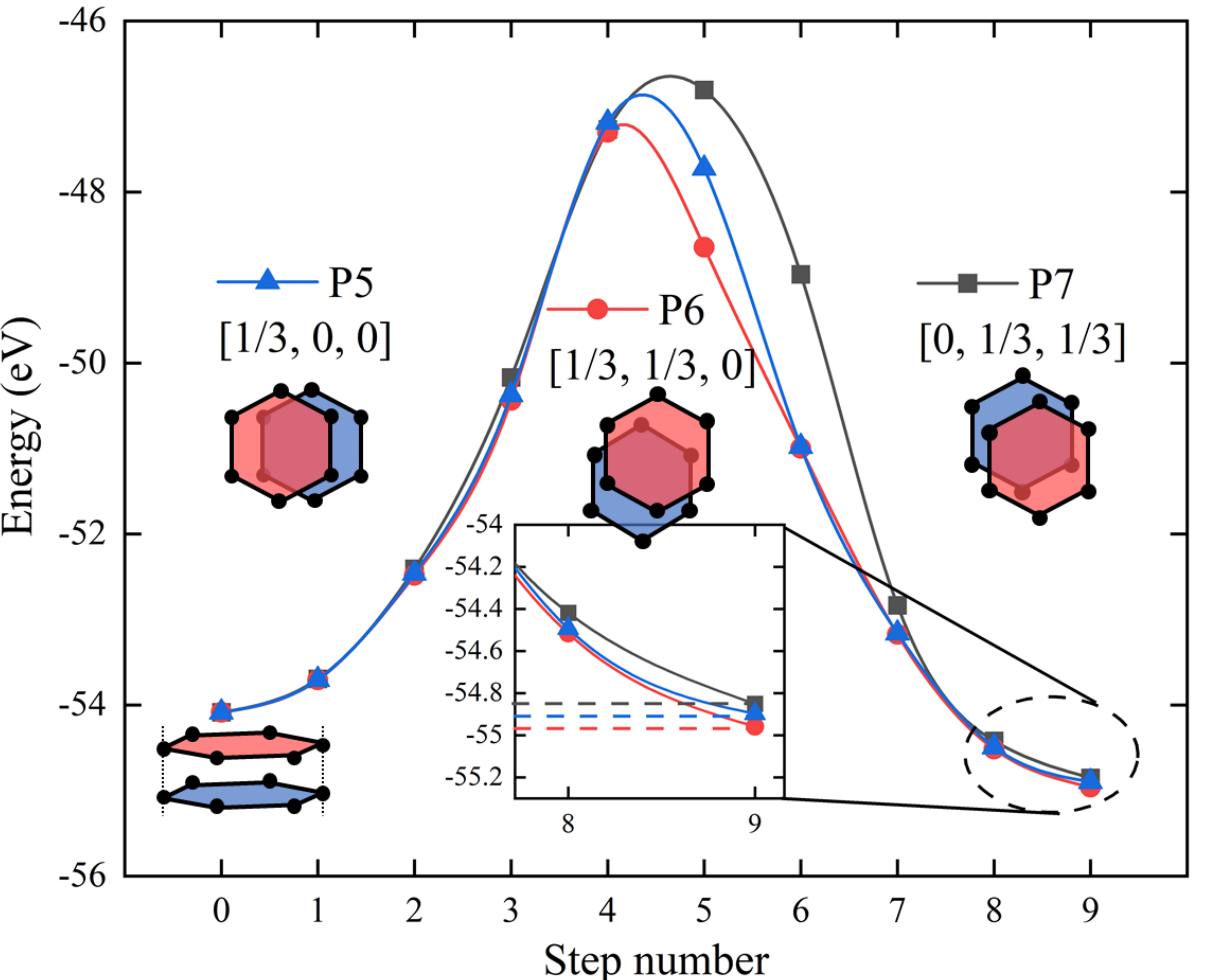}
 \caption{\label{3} (a) The transition path and the activation energy per primitive cell from initial P1 to final states P5, P6 and P7, respectively, as a function of step number. The inset is zoom-in figure for dotted oval area. Red hexagon denotes V hexagons while blue denotes Cr hexagons. Note that I atoms are hidden in the schematics.}
     \end{figure}

Then, we set P1 as initial state while P5 $\sim$ P7 as final states to investigate the potential barriers among them. Seven transient states onto the reaction paths are linearly interpolated. The evolution of reaction energy shows that the final states are 0.77 $\sim$ 0.87 eV/primitive cell lower than initial P1 (Fig. \ref{3}). Among them, P6 is the most energetically favorable. It also shows large potential barriers all in the three cases as 7.17 eV, 6.31 eV and 8.04 eV/primitive cell for P5, P6 and P7, respectively. It can be partly attributed to: (i) the movement of V layer in P1 should overcome exchange interaction from the sublayer Cr atoms . The nearest-neighboring Cr atom is only one for each V atom with a distance $d_{V-Cr}$ = 4.183 {\AA}. It is boosting before moving to final states whose nearest-neighboring Cr atoms are doubled ($d_{V-Cr}$ = 4.381 {\AA}); (ii) the octahedron units formed by V-I$_6$ are destroyed during the moving process, which results in elevated energy; (iii) the movement of top I layer should also overcome interaction from sublayer Cr atoms. Anyway, such large potential barrier is desired because it hinders transition between P1 and P6 due to moderate thermal effect, strain induced from substrate or other external factors.

\begin{figure}[htbp]
\centering
\includegraphics[scale=0.47]{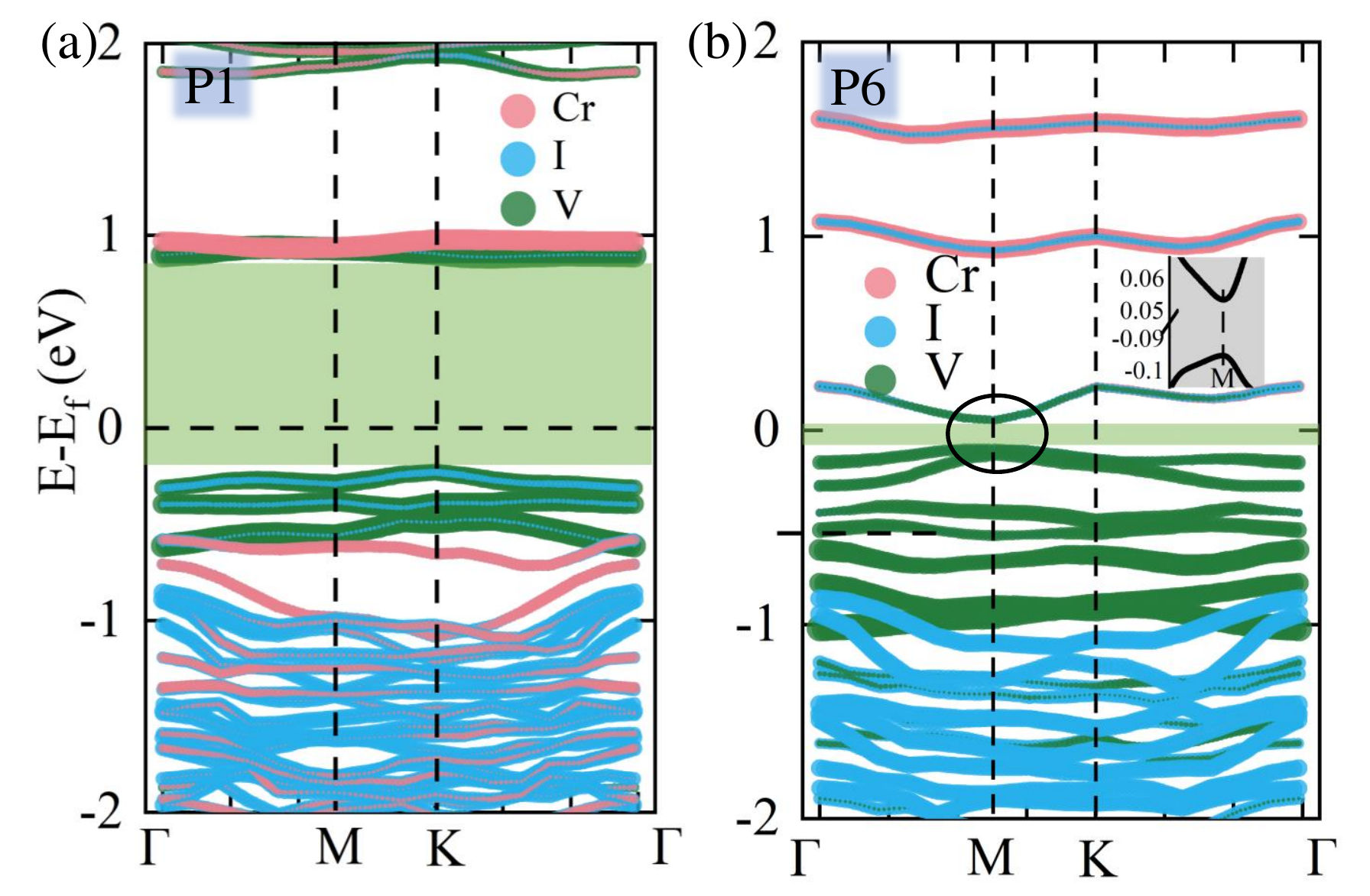}
\caption{\label{4}(a, b) Projected band structures on Cr, I and V atoms for P1 and P6 of monolayer V$_2$Cr$_2$I$_9$, respectively.  The insert in (b) shows the zoom-in figure for black oval area. The green area in (a) and (b) denote band gaps.}
\end{figure}

%\subsection{Both Weakened Direct and Indirect Exchange Interaction as Stretching}
The band structure of P6 shows a $\sim$ 0.15 eV direct band gap whose valence band maximum (VBM) and conduction band minimum (CBM) both occur at M point [Fig. \ref{4}(b)]. Relatively, in P1 it shows an indirect type but with a larger gap of 1.15 eV. In both cases, I atoms mainly occupy lower states less than -1.23 eV, while states of V atoms locate mainly near the fermi level. The main discrepancy is, Cr atoms take up only empty state and are highly localized at $\sim$ 1.00 eV and $\sim$ 1.56 eV in P6 [Fig. \ref{4}(b)], while in P1, they occupy fully filled band states from -1.4 eV to -2 eV [Fig. \ref{4}(a)], indicating a strong hybridization between Cr and I atoms. Meanwhile, the VBM and CBM mainly originate from V and I atoms in P1 while from I atoms in P6.

%\subsection{Enhanced Ferromagnetism in VXY}

\begin{figure}[htbp]
    \centering
    \includegraphics[scale=0.5]{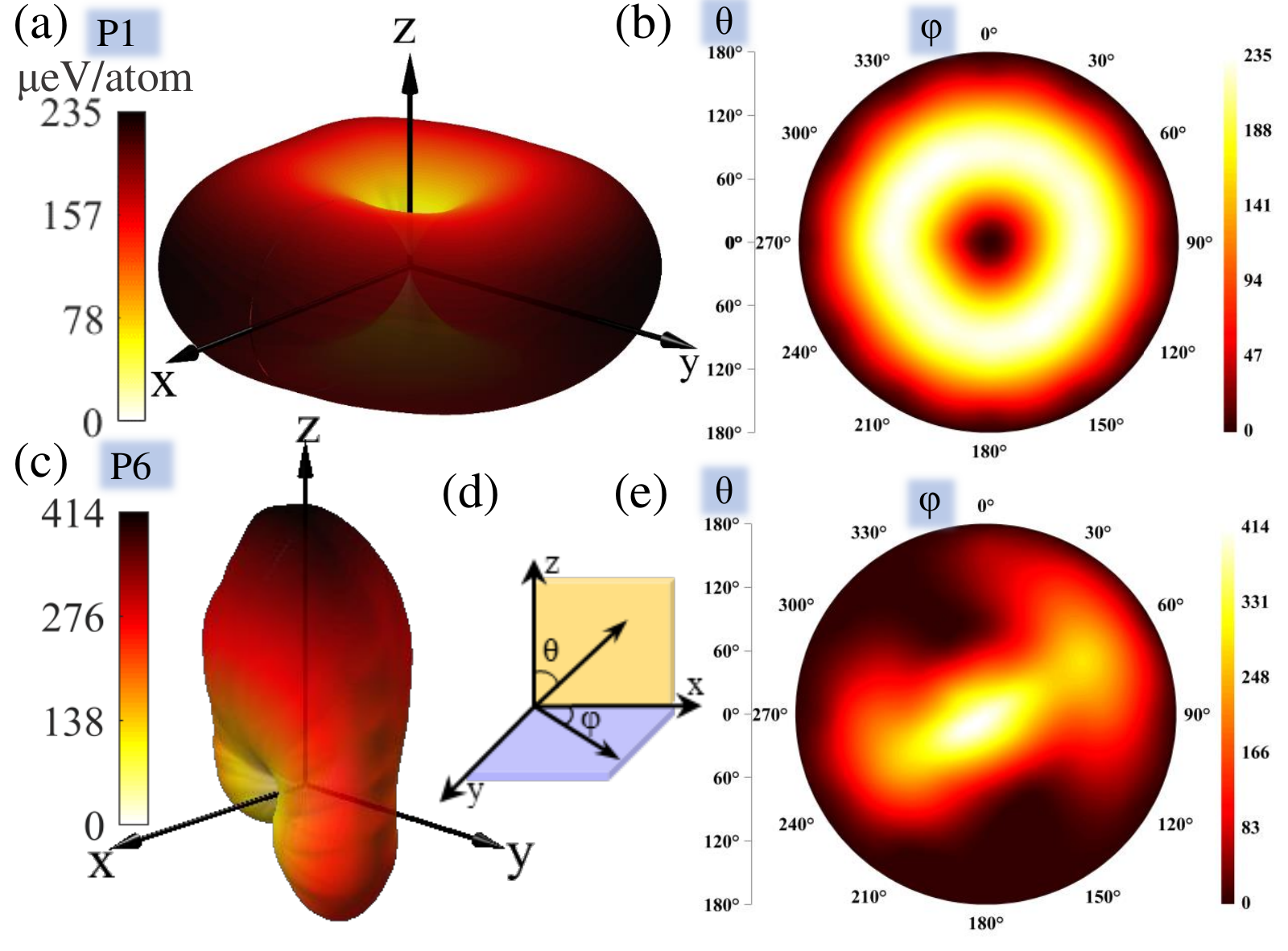}
       \caption{\label{5}(a, c) Angular dependence of MAE with the direction of magnetization lying on the whole space for P1 and P6, respectively. (b, e) The whole-space MAE projected onto polar coordinates for P1 and P6, respectively. The vertical coordinates denote $\theta$ and circle coordinates denote $\varphi$. (d) The spin vector S on the x, y, and z planes rotated with angles out-of-plane $\theta$ and in-plane $\varphi$. Note that heavier coloration indicates the increase in MAE in (a, c) and inversely, heavier coloration indicates the decrease in (b, e). The calculation of MAE takes spin-orbital coupling (SOC) effect into consideration.}
     \end{figure}

The magnetic anisotropy energy (MAE) is considered as a requisite in 2D materials according to Mermin-Wagner theorem \cite{Mermin1966}. To investigate the ferromagnetism that exists in V$_2$Cr$_2$I$_9$ monolayer, we calculate the angular-dependent MAE for P1 and P6 (Fig. \ref{5}), defined as MAE=$E_{\theta,\varphi} -E_{0, 180}$, where $\theta$ and $\varphi$ denote polar angle and azimuth angle in polar coordinate system [Fig. \ref{5}(d)]. In P1, it displays that heavy coloration basically locates at the boundary of the cookie-like MAE distribution as shown in Fig. \ref{5}(a). Meanwhile, the ring-shaped distribution means MAE remains unchanged when spin vector S switches from $\varphi = 0^{\circ}$ to $\varphi = 360^{\circ}$ in ab plane [Fig. \ref{5}(b)]. But when it rotates from $\theta = 0^{\circ}$ to $\theta = 90^{\circ}$, a clear out-of-plane anisotropy is demonstrated with the MAE elevating from 0 to 234.7 $\mu$eV per Cr (V) atom. In P6, it shows a lower symmetry with MAE concentrating on [0 0 1] direction [Fig. \ref{5}(c)]. Furthermore, there exist not only in-plane anisotropy but also out-of-plane anisotropy [Fig. \ref{5}(e)]. When $\theta$ increases from 0 to $180^{\circ}$, the MAE declines from maximum as 412.9 $\mu$eV per Cr (V) atom in [0 0 1] to 0 in [0 0 $\overline{1}$] direction. Therefore, the out-of-plane MAE possesses major contribution to anisotropy in P6. Additionally, the MAE declines with distinct tendency when $\theta$ increases with different $\varphi$. It probably results from the breaking rotational symmetry by in-plane shift of V layer of P1.

%\subsection{Magnetic Anisotropy and Curie Temperature}

\begin{figure*}[htbp]
    \centering
    \includegraphics[scale=0.51]{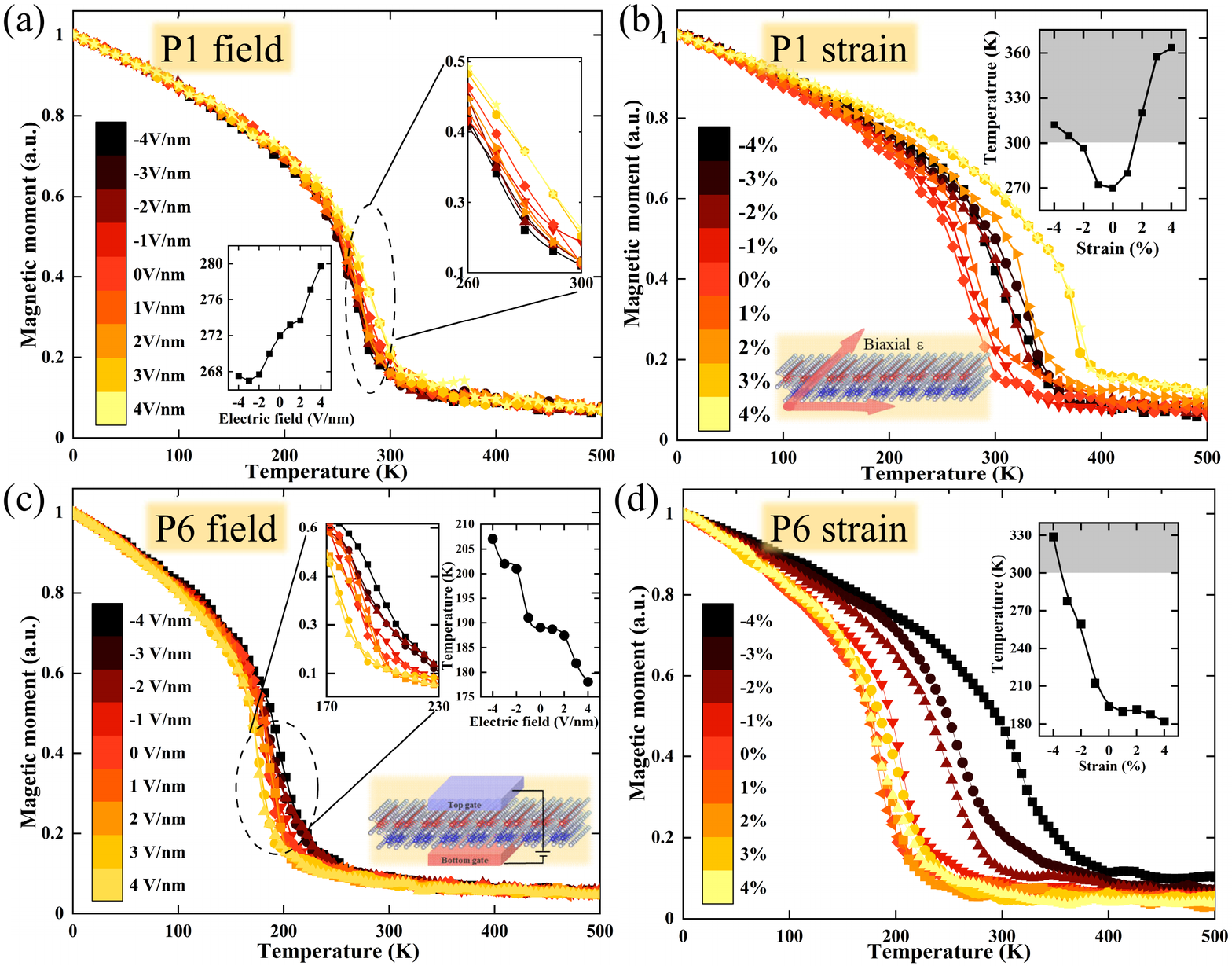}
       \caption{\label{6}(a, c) Normalized magnetization as a function of temperature under electric field ranging from -4 V/nm to 4 V/nm for P1 and P6, respectively. The insets with colored curves are zoom-in figures for dashed oval area. While the single-black curve inserts display $T_c$ as a function of electric field. Additionally, at the bottom panel of (c) it shows a schematic of a 2D magnetoelectric device with a longitudinal electric field. (b, d) Magnetization as a function of biaxial strain from -4 \% to 4 \% for P1 and P6, respectively. Similarly, the inserts show $T_c$ as a function of strain. Note that the results are all calculated with SOC.}
     \end{figure*}

We estimate $T_c$ by using Monte Carlo simulation with Heisenberg model \cite{supply}. The in-plane nearest-neighboring exchange constant J$_{1\bot}$, next-nearest-neighboring exchange constant J$_{2\bot}$ and out-of-plane nearest-neighboring exchange constant J$_{1\|}$ are considered. J$_{1\bot}$ is estimated to be -25.30 (-16.93) meV for P1 (P6), which is 8 (5) times higher than monolayer CrI$_3$. According to the results, $T_c$ for ground-state P1 (P6) is calculated as $\sim$ 272 K ($\sim$ 194 K, Fig. \ref{6}). It shows a significant enhancement compared to the prototypes e.g. monolayer CrI$_3$ and VI$_3$. The reason for boosted $T_c$ can be mainly attributed to: (i) large MAE per Cr (V) atom, which means more thermal energy is requisite to disorder long-range spin arrays. So the energy difference between antiferromagnetism and ferromagnetism configurations is expanded; (ii) The amount of near-90$^{\circ}$ I-Cr(V)-I bond angles per primitive cell is doubled from CrI$_3$ and VI$_3$ compounds to V$_2$Cr$_2$I$_9$ monolayer, leading to an enhancement of superexchange interaction and a more stabilized parallel spin configuration \cite{Goodenough1955, Goodenough1958, Kanamori1960, Anderson1959}. Although direct exchange interaction is meanwhile moderately enhanced, it is more sensitive to the distance between V-V (Cr) than superexchange interaction \cite{Ren2020}.

Inspired by recent works on gate-dependence CrI$_3$, we also employ experimentally accessible longitudinal electric field of a few V/nm for P1 and P6, as displayed in Figs. \ref{6}(a) and \ref{6}(c). Unlike quite narrow tunable range of $T_c$ in P1 (i.e. from 266 K to 280 K), it shows a moderate change range of $T_c$ in P6: When the electric field is tuned from positive to negative values, it enables a variation of $T_c$ between 177 K and 208 K. Besides, $T_c$ is elevated by switching electric field from negative to positive in P1 while in P6 it declines.

Strain effect ($\varepsilon$), directly affects the bond lengths which possess significant influence on exchange interaction in d-orbital honeycomb crystal. Therefore, the strain effect on monolayer V$_2$Cr$_2$I$_9$ is investigated as shown in Figs. \ref{6}(b) and \ref{6}(d). A rather different ferromagnetism tuning behaviors are found in P1 and P6. In the former, tensile and compressive strain both enable an enhancement of $T_c$, which is more sensitive to tensile strain. In the latter, $T_c$ is drastically elevated from 193 K to 328 K as $\varepsilon$ changing from 0 to $-4 \%$. But for $0\leq \varepsilon\leq 4 \%$, $T_c$ remains just a slightly decrease from 193 K to 182 K. In addition, it is rather fascinating that $T_c$ surpasses room temperature when $|\varepsilon|$  $\geq$ 2 \% in P1 and $\varepsilon = -4 \%$ in P6.
\section{Conclusion}
By using first-principles calculations, we first combine the monolayer VI$_3$ and CrI$_3$ into the newly constructed bimetal transition iodide V$_2$Cr$_2$I$_9$ monolayer and investigate the stability for a set of possible phases, which are obtained by switching the stacking order of I atoms and by giving a rigid shift to V-I$_3$ layer to [1/3 0 0], [1/3 1/3 0] and [0 1/3 0]. Among the 7 phases, it is determined that only P6 is stable while P1 is metastable. By band structure calculations, it shows a larger band gap for P1 while direct-type band gap for P6. Next, crucial for the ferromagnetism exhibiting in dimensionality $\leq$ 2, is the MAE estimated as a large value especially in P6. After that, we have discussed that $T_c$ is 6 (4) times larger than monolayer CrI$_3$ and VI$_3$ in ground state P1 (P6).  Furthermore, $T_c$ is highly dependent on biaxial electric field and strain, we can achieve an alteration of $T_c$ by switching direction of electric field and by employing external strain. It is believed that this work supports a new idea to combine two transition metal elements into bimetal transition iodide monolayer by structure engineering. Meanwhile, monolayer V$_2$Cr$_2$I$_9$ probably possesses potential application in spintronic devices as FMSs with the $T_c$ near room temperature.

\section{METHODS}
The first-principles calculations were performed using the projected augmented-wave method \cite{Kresse1999, Blochl1994} as implemented in the Vienna Ab initio Simulation Package (VASP) \cite{Kresse1996} based on Perdew-Burke-Ernzerhof (PBE) functional. In our global calculations, we used a 450 eV cut-off energy, 0.05 eV width of smearing and a $7\times7\times1$ grid of the first Brillouin-zone integration. Also, we remained the thickness of vacuum spacing unchanged as 20 {\AA} to reduce the interlayer interactions due to the periodic boundary conditions. In most parts of the calculations, the Hellman-Feynman force and total energy convergence criteria were employed less than $10^{-2}$ meV/{\AA} and $10^{-6}$ eV. Note that the phonon-frequency and MAE calculations require rather high precision, so the two convergence criteria were set to $10^{-3}$ meV/{\AA} and $10^{-8}$ eV to suffice the high accuracy of the results. With regard to the crystal volume, a $3\times3\times1$ supercell model (117 atoms in total) was applied to perform phonon-frequency calculations and a $2\times2\times1$ supercell model (52 atoms in total) in molecular-dynamic simulations. Also, a $2\times2\times1$ supercell model was used to simulate different spin configurations (e.g. N\'{e}el antiferromagnetic configuration). In addition, the band structures were given by Heyd-Scuseria-Ernzerhof (HSE06) functional including 25\% non-local Hartree-Fock exchange \cite{Heyd2003}. The non-collinear calculations were given by considering SOC effect \cite{SOC} in MAE and electric field simulations. The phonon frequency calculations were carried out by using DFT perturbation theory as implemented in the PHONOPY code \cite{Togo2008}. And the AIMD simulations in the canonical (NVT) ensemble were performed at 300 K with a Nos\'{e} thermostat.

\begin{acknowledgments}
This work was supported by National Natural Science Foundation of China (No.11904312 and 11904313), the Project of Hebei Education Department, China (No.ZD2018015 and QN2018012), and the Natural Science Foundation of Hebei Province (No.A2019203507). Thanks to the High Performance Computing Center of Yanshan University.
\end{acknowledgments}

\bibliography{apssamp}% Produces the bibliography via BibTeX.

\begin{thebibliography}{99}
\bibitem{supply} Supporting material is available that contains details for crystal information, phonon dispersion relations for the other phases, estimation of $T_c$ and so on.
\end{thebibliography}

\end{document}